# Joint Modelling of Dust Scattering and Thermal Emission: The Spider Complex

Jielai Zhang (张洁莱)[1,2,3,4,5], Peter G. Martin[5], Ryan Cloutier[1,6], Natalie Price-Jones[1], Roberto Abraham[1,2], Pieter van Dokkum[7], and Allison Merritt[7,8]
[1] Department of Astronomy and Astrophysics, University of Toronto, 50 St. George Street, Toronto, ON M5S 3H4, Canada; jzhang@schmidtsciencefellows.org
[2] Dunlap Institute, University of Toronto, 50 St. George Street, Toronto, ON M5S 3H4, Canada
[3] Centre for Astrophysics and Supercomputing, Swinburne University of Technology, Hawthorn, VIC, 3122, Australia
[4] Australian Research Council Centre of Excellence for Gravitational Wave Discovery (OzGrav), Australia
[5] Canadian Institute for Theoretical Astrophysics, University of Toronto, 60 St. George Street, Toronto, ON M5S 3H8, Canada
[6] Dept. of Physics & Astronomy, McMaster University, 1280 Main St West, Hamilton, ON L8S 4L8, Canada
[7] Department of Astronomy, Yale University, 52 Hillhouse Avenue, New Haven, CT 06511, USA
[8] Max-Planck-Institut fur Astronomie, Konigstuhl 17, D-69117 Heidelberg, Germany
Received 2022 October 21; revised 2023 March 3; accepted 2023 March 3; published 2023 April 27

## Abstract

Observations across the electromagnetic spectrum of radiative processes involving interstellar dust—emission, absorption, and scattering—are used to constrain the parameters of dust models and more directly to aid in foreground removal of dust for extragalactic and cosmological observations. Dust models can benefit from more independent constraints from complementary observations. Here, we quantify the relationship between scattered light and thermal emission from dust in a diffuse (cirrus) intermediate-latitude cloud, Spider, using data from the Dragonfly Telephoto Array and the Herschel Space Observatory. A challenge for optical observations of faint diffuse cirrus is accurate removal of a contaminating, spatially varying sky. We present a technique to analyze two images of the same cirrus field concurrently, correlating pixel values to capture the relationship and simultaneously fitting the sky-related signal as a complex noncorrelating additive component. For the Spider, we measure a color $g - r = 0.644 \pm 0.024$ and ratios of visible-wavelength to 250 $\mu$m intensity of $\gamma_{g,250} = (0.855 \pm 0.025) \times 10^{-3}$ and $\gamma_{r,250} = (1.55 \pm 0.08) \times 10^{-3}$ for the $g$ and $r$-bands, respectively. We show how to use any dust model that matches the thermal dust emission to predict an upper limit to the amount of scattered light. The actual brightness of the cirrus will be fainter than this limit because of anisotropic scattering by the dust combined with anisotropy of the incident interstellar radiation field (ISRF). Using models of dust and the ISRF in the literature, we illustrate that the predicted brightness is indeed lower, though not as faint as the observations indicate.

*Unified Astronomy Thesaurus concepts:* Interstellar dust (836); Interstellar medium (847); Astronomical techniques (1684); Submillimeter astronomy (1647); Optical astronomy (1776); Interstellar scattering (854); Dust continuum emission (412); Astronomy image processing (2306)

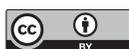

## 1. Introduction

Dust is a critical component of the interstellar medium (ISM) and through many physical processes it impacts both observational astronomy and galactic evolution (Martin 1978; Draine 2011). For extragalactic astronomy and cosmology, Galactic dust is an unwanted foreground in extinction (Barnard 1907, 1910; Trumpler 1930; Schlafly & Finkbeiner 2011) and polarized thermal emission (e.g., BICEP2/Keck Array & Planck Collaborations 2015). Understanding the physical and radiative properties of dust has therefore been of longstanding interest and yet it is still unsettled and so an active area of investigation. This paper shows how joint modeling of dust scattering and thermal emission can contribute.

Dust models used to predict absorption, scattering, and emission properties require knowledge of the shapes, size distributions, compositions, and dielectric functions of the grains. These are constrained by a wide range of astronomical observations of (polarized) extinction and (polarized) emission combined with available elemental abundances (Hensley & Draine 2021). There have been many Galactic dust models based on separate populations of amorphous silicate and carbonaceous grains (Weingartner & Draine 2001; Draine & Li 2007; Draine & Fraisse 2009; Compiègne et al. 2011), but with the advent of submillimeter observations of (polarized) thermal emission these models have been found wanting and in need of improvement (Hensley & Draine 2021). One approach is to have a single population of dust grains incorporating both compositions, as motivated and developed by Draine & Hensley (2021) and Hensley & Draine (2022).

While extinction and emission have been used extensively to inform dust models, scattered light (familiar in reflection nebulae) has been used less because the amount of scattered light is dependent on not only the scattering cross section, but also the scattering phase function, the direction of the incident radiation, and optical depth effects. Diffuse Galactic light (DGL) offers perhaps the best constraints, but it is difficult to observe and complex to model (Hensley & Draine 2021). Extensive observations from the near-infrared to ultraviolet obtained by correlation with 100 $\mu$m dust emission from IRAS are discussed in Section 6.

### 1.1. Integrated Flux Nebulae

Sandage (1976) identified faint structures in images taken at high Galactic latitude that were interpreted as scattered light nebulosities illuminated by the Galactic plane. These have been splendid subjects for deep imaging by dedicated, talented amateur astronomers, who call them integrated flux nebulae (IFNs) because they are illuminated not by the light of a single star or cluster but by the interstellar radiation field (ISRF) of the Galaxy. Targeted scattered light observations to quantify the brightness of such diffuse patches of optically thin cirrus are difficult because of their faintness, typically only a few percent of the brightness of the dark night sky, but offer new opportunities.

Pioneering observations by Witt et al. (2008) measured the surface brightness of scattered light in *B*-band of five optically thin high Galactic latitude diffuse clouds (IFNs) and found it to be a factor of 2–3 fainter than expected for an ISRF strength given by Porter & Strong (2005) if the scattering were isotropic, pointing to the importance of modeling the anisotropy of both the scattering phase function and the ISRF. They also found a strong correlation of the scattered light with the 100 $\mu$m IRAS brightness for the optically thin clouds. Ienaka et al. (2013) compared optical scattered light (in *B*, *g*, *V*, and *R* bands) with 100 $\mu$m emission from cloud MBM32 and found that differences in optical depth can change the ratio of scattered light to thermal emission by a factor of 2–3, for $A_V$ in the range 0.16–2.0. Reprocessing Sloan Digital Sky Survey (SDSS) imaging data optimized for low-surface-brightness sources in the Stripe82 region, Román et al. (2020) found that by using $r - i$ and $g - r$ colors cirrus can be differentiated from extragalactic sources. They also found that the colors of cirrus are dependent on dust column density, again pointing to the simplification that comes with optically thin clouds.

### 1.2. Our Focus

The Spider complex (Spider for short) was chosen for this work because a large fraction of the cloud is optically thin and it has been mapped in submillimeter thermal dust emission using the Herschel Space Observatory[9] (hereafter Herschel). The Spider field is centered on $\ell \sim 135°$, $b \sim 40°$ and is at a distance of about 320 pc (Zucker et al. 2020; Marchal & Martin 2023), thus at a height $z = 205$ pc above the Galactic plane. It is outside the Local Bubble (Zucker et al. 2021) but appears to be related to a extended protrusion (Marchal & Martin 2023). For broader context, the Spider field is at the top of the arch of the North Celestial Pole Loop (NCPL, Martin et al. 2015; Blagrave et al. 2017; Taank et al. 2022; Marchal & Martin 2023). Areas in the NCPL have been the target of many IFN images, such as one by R. B. Andrea[10] that includes what we call the Spider.

Our scattered light images of Spider come from the Dragonfly Telephoto Array (hereafter Dragonfly; Abraham & van Dokkum 2014), which is optimized to image large-area extended sources with low surface brightness (Section 2.1). The main orientation of Dragonfly science is extragalactic, but optically thin cirrus can be measured quantitatively too (often as an unwanted contaminant). Dragonfly imaging in the Sloan *g* and *r* bands can characterize the color of the scattered light (Section 3).

Herschel imaging of thermal dust emission offers an increase in resolution by a factor of about 15 (at 250 $\mu$m) compared to the IRAS 100 $\mu$m maps and yet is very sensitive (Section 2.2). Combined, Dragonfly and Herschel can determine the ratio of intensities of scattered light and thermal emission from dust (Section 4). Joint modeling of this ratio provides a useful new quantitative constraint on models of dust and the ISRF (Section 5).

The relationship to observations and modeling of the DGL is discussed briefly in Section 6. Our summary and conclusions are in Section 7.

## 2. Data

### 2.1. Scattered Light

#### 2.1.1. Dragonfly Imaging

Scattered light observations were acquired using Dragonfly (Abraham & van Dokkum 2014), a multilens visible-wavelength telescope optimized for low-surface-brightness observations. At the time of the observations used, Dragonfly consisted of 48 imaging units. Each unit was a Canon 400/2.8 L IS II USM lens, a thermoelectrically cooled science grade CCD by Santa Barbara Imaging Group (SBIG), and a Kodak KAF-8300 sensor (models ST, SFT, or STT). The pixel size was 2″.85. Half of the lenses were equipped with Sloan *g* filters ($\lambda_0 \sim 0.48$ $\mu$m) and the other half with Sloan *r* filters ($\lambda_0 \sim 0.62$ $\mu$m). The field of view of each lens is 2°.6 by 1°.9 and Dragonfly observed with all lenses pointing in the same direction (not offset by more than 10′). The 48 units were arranged on two separate Paramount Taurus mounts, manufactured by Software Bisque Inc., each holding 24. The different footprints on the sky for *g*- and *r*-band images in Figure 1 are a result of stacking images from cameras with nonidentical offset directions. The Dragonfly Arrays are situated at the New Mexico Skies Observatory in Cloudcroft, New Mexico, U.S.A.

Dragonfly images used in this paper were obtained during two epochs: 2014 and 2017. While 244 and 258 images, each with an exposure of 10 minutes, were taken in the two epochs, respectively, not all images were usable. Images that were affected by thin atmospheric clouds, were affected by stray sources of light, or were out of focus were not used in the analysis. Notably, many images taken in 2014 were out of focus, having been taken during the development of new automated focusing scripts for the telescopes. The numbers of exposures used in the final analysis by filter and epoch are summarized in Table 1. The combined Dragonfly *g*- and *r*-band images taken in 2017 are shown in Figure 1.

#### 2.1.2. Dragonfly Data Reduction

The Dragonfly data reduction pipeline (Zhang 2021) optimized for low-surface-brightness extragalactic science was not used verbatim to process Dragonfly data in this study. The sky removal procedure in that pipeline removes signals on scales larger than about a third of the field of view by fitting a 2D polynomial to sky regions and subtracting this additive component. Cirrus data used in this study extend to scales covering the entire image and background regions cannot be identified reliably. We provide a brief description of the main data reduction steps that follow the Dragonfly data reduction pipeline and describe customizations made for the cirrus data studied in this paper.

The Dragonfly *g*- and *r*-band images were first dark-subtracted and flat-fielded. Each imaging unit has its own set of dark and flat-field images to do this calibration. A world coordinate system (WCS) was embedded using `Astrometry.net` (Lang et al. 2010). Image registration was performed using the `Astromatic.net` packages Source Extractor (Bertin & Arnouts 1996), SCAMP (Bertin 2006), and SWarp (Bertin et al. 2002). Source Extractor and SCAMP were used to measure the astrometric distortions of each image by comparing source positions in the Dragonfly images to an online astrometric catalog. SWarp was used to resample and register all images to a common grid. After image registration, the nonlinear distortions are smaller than 1″ across the entire image. During the resampling process, the pixel scale of each Dragonfly exposure was changed from 2″.85 pixel$^{-1}$ to 2″.0 pixel$^{-1}$, and pixel values were lowered by a factor of $2.85^2/2.0^2 = 2.03$. This means that the noise associated with a pixel value of $X$ in the final combined image actually had the amount of Poisson noise with a pixel value of $2.03X$ ADU. This is equivalent to a change in gain when calculating Poisson noise. The effective gain after registration of images is $0.37 \times 2.03 = 0.75$, where 0.37 was the original gain of the CCDs.

---

[9] Herschel is an ESA space observatory with science instruments provided by European-led Principal Investigator consortia and with important participation from NASA.

[10] https://www.deepskycolors.com/archive/2010/04/08/integrated-Flux-Nebula-Ifn-really-wide.html

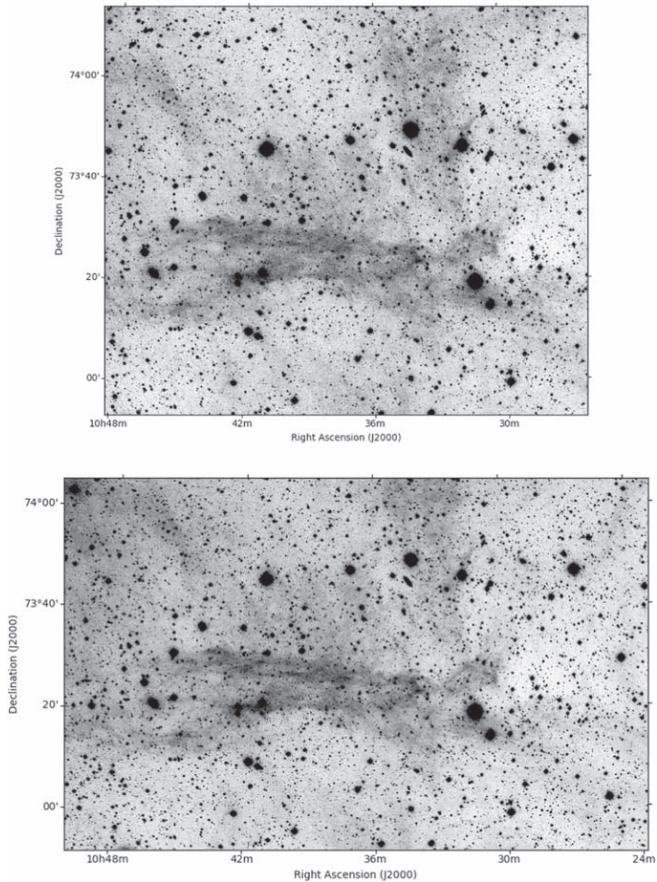

**Figure 1.** Reduced and stacked *g*-band (upper) and *r*-band (lower) images of Spider taken with the Dragonfly Telephoto Array in 2017.

**Table 1**
Summary of Dragonfly Observations: Number of 10 Minutes Exposures Stacked and FWHM

| Epoch | Sloan filter | $\lambda_0$ ($\mu$m) | Number | FWHM |
|---|---|---|---|---|
| 2014 | g | 0.48 | 46 | 6″.4 |
|  | r | 0.62 | 21 | 6″.6 |
| 2017 | g | 0.48 | 93 | 6″.8 |
|  | r | 0.62 | 117 | 8″.0 |

Individual resampled images were visually inspected to remove frames with non-photometric conditions, large sky gradients, and obscuration by atmospheric clouds. We then summed these registered images separately for *g*- and *r*-band images. A zero-point was calculated for the stacked images by comparison to the APASS catalogs (Henden et al. 2016). The photometric errors of the stacked Dragonfly images are $<\pm 0.06$ mag. The sum operation was used here to because it allows a more direct calculation of the Poisson noise of the combined Dragonfly image. Also, to assist in calculating Poisson noise, the flat-fielding procedure was performed with flat frames normalized so that the central pixels retain the native ADU unit of raw frames. As a result of using the sum operation, satellite trails remain in the stacked images and were masked in subsequent analysis. Table 1 shows the average FWHM of point sources in the stacked images.

To make the correlation slopes discussed below more directly physically meaningful we converted the Dragonfly image pixel values to surface brightness $I_D$ in kJy sr$^{-1}$ using Equation (1):

$$I_D[\text{kJy sr}^{-1}] = \frac{\text{ADU}}{10^{\text{ZP}/2.5}} \times \frac{3631}{1000} \times \frac{(180 \times 3600/\pi)^2}{4}, \quad (1)$$

where ADU is the pipeline value in the coadded image with 2″ pixels, and ZP is the zero-point measured for the image.

### 2.2. Thermal Emission

To measure the thermal emission from dust, images from the Spectral and Photometric Imaging Receiver (SPIRE) instrument on Herschel were used (Griffin et al. 2010). SPIRE performed imaging in three bands, centered on 250, 350, and 500 $\mu$m. From the online Herschel Science Archive (HSA),[11] we obtained the high-quality Level 2.5 products created using HIPE[12] pipeline version 14.0. All Level 2.5 products combine scan and cross-scanned data optimally in a single map at each frequency. The SPIRE images have had a zero-level correction and pixel values are in MJy sr$^{-1}$. In fact, all SPIRE and PACS images can be put on an absolute scale (zero level corrected: Singh & Martin 2022) by correlation with the Planck all-sky thermal dust model, in which the cosmic infrared background monopole has been eliminated (Planck Collaboration XI 2014).

The 250 $\mu$m band was chosen because it has the highest signal-to-noise ratio. It has a beam FWHM of 18″, so that the resolution is substantially lower than Dragonfly's. The 250 $\mu$m image of Spider is shown in Figure 2. The regions that correspond to each of the four Dragonfly reduced images, as listed in Table 1, are indicated with boxes.

## 3. Scattered Light Color

### 3.1. Source Masking

To measure the $g - r$ color of the scattered light from Spider, stars and galaxies must first be masked. One option is to generate a mask using the segmentation map output by applying Source Extractor to each combined image. This segmentation map identifies groups of connected pixels that exceed a threshold above the background. A second option is to simulate the stars and galaxies in the combined Dragonfly image in each filter band and then create a mask using a threshold in each simulated image. Both masking methods will unavoidably leave light from sources beyond their bright central regions unmasked. Knowing this, and anticipating convolution to the Herschel resolution too, we chose the second option. Another advantage of this option is that a measure of the contribution of noise due to unmasked light from stars and galaxies can be made. The more accurate the simulated image, the better the estimate of this source of error.

---

[11] http://archives.esac.esa.int/hsa/whsa/
[12] HIPE is a joint development software by the Herschel Science Ground Segment Consortium, consisting of ESA, the NASA Herschel Science Center, and the HIFI, PACS, and SPIRE consortia.

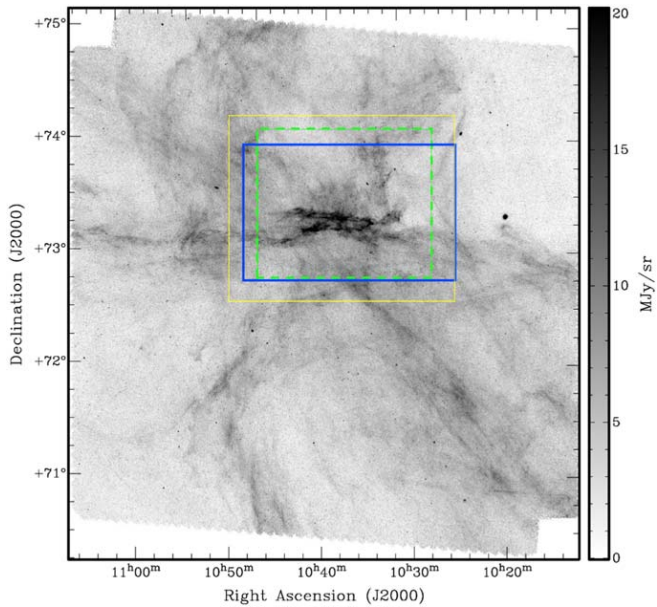

**Figure 2.** Herschel 250 μm image of the Spider. The largest (thin, yellow outline) box shows the footprint of the Dragonfly g- and r-band combined images from 2014. The most elongated (bold, blue solid outline) box shows the footprint of the Dragonfly r-band combined image from 2017. The smallest (dashed, green outline) box shows the footprint of the Dragonfly g-band combined image from 2017.

The Astromatic.net software package SkyMaker (Bertin 2009) was used to simulate the Dragonfly stacked images. SkyMaker requires an input catalog, which we created by applying Source Extractor to the stacked Dragonfly images. The magnitudes of sources were cross-matched with the APASS catalog (Henden et al. 2016) as a sanity check. In the magnitude range where APASS photometry is reliable (sources brighter than about 17.5 mag in both bands), the APASS magnitude was used, although this would not have made a big difference to the catalog to be simulated because the photometry error on the Dragonfly images is $<\pm 0.06$ mag.

The SkyMaker internal point-spread function (PSF) generator models the effects of atmospheric blurring and instrument diffraction and aberrations. The seeing FWHM, number of pupil obscuring arms (zero in our case) and primary lens diameter are input into SkyMaker to generate the PSF.[13] On inspection it appears that some of the faint sources in the Source Extractor catalog were simply knots of bright cirrus, making the mask conservative.

---

[13] More sophisticated wide-angle PSFs can be applied and tested in this step but fall outside the scope of this paper, e.g., Liu et al. (2022).

The simulated Dragonfly images were then convolved to 18″ (or 36″) to match the Herschel resolution and resampled onto the Herschel grid using SciPy4. The threshold used to turn the simulated image into a mask was chosen to mask regions of stars and galaxies with signal above the Poisson fluctuations of the sky.

### 3.2. Sky Subtraction

Sky subtraction is a central problem affecting low-surface-brightness photometry. Sandin (2015) showed that over- and undersubtraction of the sky can easily affect the photometry and the structure of measured low-surface-brightness sources that have only a fraction of the sky brightness. Numerous studies have wrestled with the difficulty of sky subtraction, including Adelman-McCarthy et al. (2008), Hyde (2009), Hyde & Bernardi (2009), Blanton et al. (2011), Zhang (2021), and Borlaff et al. (2019). Techniques used include improved flat-fielding and iterative sky fitting with increasingly better extended source masking. These techniques, however, cannot be applied to cirrus data, as cirrus fills the entire field of view of acquired images, and no sky pixels can be reliably identified.

We present a novel technique of simultaneously correlating multiband cirrus data and accounting for the sky. Dragonfly diffuse imaging data over pixel $(x, y)$ positions, $D(x, y)$, are a combination the scattered light intensity, $I(x, y)$, plus the sky component, $S(x, y)$. The premise for our technique is that $I_g(x, y)$ and $I_r(x, y)$ for the two bands are linearly correlated for optically thin cirrus, say with a slope $q$, which is a measure of the color of the cirrus. This is propagated into the relationship between observed data so that

$$D_g(x, y) = q\, D_r(x, y) + P(x, y), \qquad (2)$$

where the additive component $P(x, y)$ is the linear combination $S_g(x, y) - q\, S_r(x, y)$ of the unknown sky components.

The $P(x, y)$ term in Equation (2) was expressed in terms of the orthogonal Zernike polynomials (Zernike 1934) with coefficients to be determined simultaneously with $q$. The use of an orthogonal basis allows the significance of adding higher-order terms to be assessed. We found that quadratic was certainly needed to model this component, cubic was justified, and some quartic polynomials could be significant, but not at the level that affected the estimate of the slope. We found a similar $P(x, y)$ and reached similar conclusions with a bivariate Legendre basis (Legendre 1782). From the correlations with submillimeter emission below (Equation (3)), the standard deviation of the sky across the image is about 2% or less, but this is still comparable to the cirrus signal and so one needs this level of modeling to separate $P(x, y)$ from the correlated cirrus variations. Note that we are interested in determining $q$ and that images of the two individual sky components are not obtained.

### 3.3. Jackknife Tests

We tested how robust the modeling and slope measurement is by performing jackknife tests. Different sets of Dragonfly data were correlated with each other, while performing simultaneous sky modeling. The measured slope, $q$ in Equation (2), should be unity for two images taken with the same filter.

The following jackknife tests were performed for g-band data and repeated for r-band data:

**Table 2**
Field-averaged $g$ to $r$ Intensity Ratio and Color

| Epoch | Method | Ratio | $g - r$ |
|---|---|---|---|
| 2014 | $g$ versus $r$[a] | 0.562 | 0.626 |
| | $g$ versus Herschel/$r$ versus Herschel[b] | 0.565 | 0.621 |
| 2017 | $g$ versus $r$ | 0.543 | 0.662 |
| | $g$ versus Herschel/$r$ versus Herschel | 0.540 | 0.669 |

**Notes.**
[a] Direct: from slope of $g$ versus $r$ from Section 3.
[b] Hybrid, via slopes of $g$ versus Herschel and $r$ versus Herschel from Table 3, Section 4.3.

1. data taken on two different nights in 2017: 2017 December 12 versus 2017 December 13.
2. 2017 data taken with two subsets of cameras.
3. 2014 versus 2017 data.

An important finding from these jackknife tests is that the slopes were very close to unity, the largest deviation being for 2017 $g$-band images taken with two subsets of cameras correlated against one another, which had a slope of 1.03, translating to an error of 0.03 mag. These tests support the idea that the modeling procedure is reliable.

The dispersion about the correlation line is a measure of the total pixel-to-pixel error in the Dragonfly data. This measure of error includes the independently estimated Poisson error, errors in sky modeling, and any other errors that we have not explicitly considered. Furthermore, when $\chi^2$ of the fit slightly exceeds one, this gives us an empirical revised effective Dragonfly uncertainty to take forward to the correlations of $g$ with $r$ and Dragonfly with Herschel data. That will ensure a more reliable slope for those correlations and result in something closer to $\chi^2 = 1$ in those fits.

### 3.4. $g - r$ Color

We calculated the $g - r$ color by first finding the slope of the pixel-to-pixel correlation between the $g$- and $r$-band Dragonfly images, while simultaneously modeling the noncorrelating additive component. We found that the correlation became nonlinear for bright cirrus regions, such that there was slightly more $r$-band light than the linear extrapolation from the low-intensity correlation. We addressed this issue by masking the bright cirrus near the center of Spider with a rectangular mask before reapplying the above procedure. Masking the central bright cirrus region in Spider has an effect on the slope measured at the 5% level. There are probably other systematic errors at the same level (see Section 4.3 and Zhang 2021).

We measured the $g - r$ color of Spider to be 0.626 and 0.662 using data from 2014 and 2017, respectively. These values are summarized in Table 2. The difference of 0.04 mag is a good estimate of the uncertainty in this procedure, including all systematic errors.

Looking ahead to Section 4.3, we calculated ratios of $g$- and $r$-band scattered light to Herschel thermal emission via correlation as well (Table 3). By taking the ratio of the ratios in the two bands, we obtain another measure of $g/r$ and the $g - r$ color. Values from this hybrid approach are also recorded in Table 2. They agree with those from the more direct procedure. Combining these results and taking into account

**Table 3**
Ratio of Scattered Light to Thermal Emission at 250 $\mu$m

| Sloan Filter | Epoch | Slope (kJy sr$^{-1}$ per MJy sr$^{-1}$) |
|---|---|---|
| $g$ | 2014 | $0.83 \pm 0.03$ |
| | 2017 | $0.88 \pm 0.07$ |
| $r$ | 2014 | $1.47 \pm 0.03$ |
| | 2017 | $1.63 \pm 0.18$ |

their variation, we derive a final estimate of $g - r$ of $0.644 \pm 0.024$.

## 4. Ratio of Scattered Light to Thermal Dust Emission

### 4.1. Source Masking

To create a mask for galaxies in the Herschel image, a threshold on a simulated image of sources in the Herschel image was used. The threshold used was selected in the same way as for the Dragonfly simulated image: the unmasked parts of galaxies should be fainter than the Poisson fluctuations in the Herschel data. To simulate the image, the SPIRE point source catalog at 250 $\mu$m was used.[14] Each catalog entry was simulated with a 2D Gaussian using catalog entries for the major and minor axis FWHM and a rotation parameter, and scaled by the catalog flux. Again, there is some indication that some compact patches of bright cirrus have been classified as sources. Masking using such sources is again conservative.

### 4.2. Correlation of Dragonfly versus Herschel 250 $\mu$m

Following the same procedure as Section 3.2, sky subtraction was performed simultaneously during the correlation of scattered light and thermal emission. We assume that—after source masking, convolving to the same resolution, and reprojecting to the same pixel grid—the signal in a Dragonfly stacked image ($D(x, y)$) is a linear combination of the Herschel image ($H(x, y)$), suitably scaled by a factor $m$, plus the additive optical sky component, $S(x, y)$ (recall from Section 2.2 that there is no submillimeter background in the Herschel image):

$$D(x, y) = m H(x, y) + S(x, y). \quad (3)$$

In Equation (3), the slope $m$ estimates the ratio of scattered light to thermal emission. In this case, the scattered light was measured in the $g$ or $r$ band and the thermal emission was measured at 250 $\mu$m. As in Section 3.2, we performed the fit by expanding $S(x, y)$ using orthogonal Zernike or bivariate Legendre polynomials. We reached the same conclusions regarding the necessary order of the polynomials and quantified the standard deviation of the sky component.

### 4.3. The Ratio

In regions of high intensity, the correlation becomes nonlinear, where empirically a higher thermal emission comes with less scattered light than expected from linear extrapolation of the low-intensity correlation (Figure 3). The nonlinear part of the correlation curve could be due to optical depth effects, changes in the scattering cross section, and/or changes in emissivity of the dust. Thus to calculate the slope, the bright

---
[14] European Space Agency, 2017, Herschel SPIRE Point Source Catalogue, Version 1.0. https://doi.org/10.5270/esa-6gfkpzh.

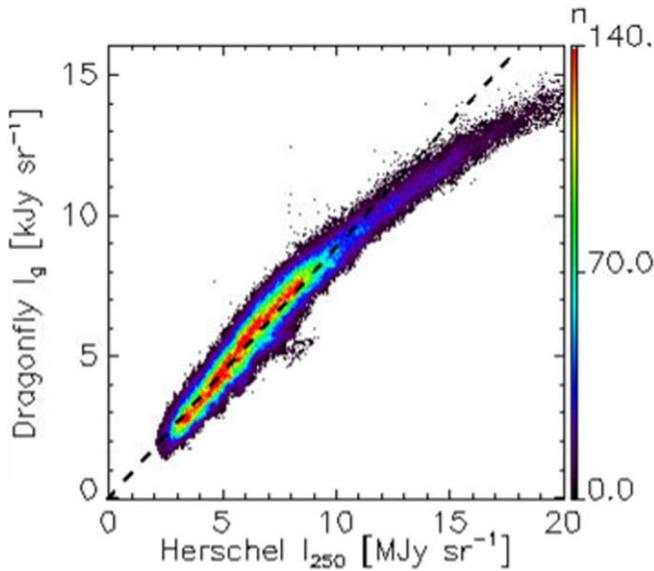

**Figure 3.** 2D histogram of intensities of scattered light (2017 $g$ band, 36″) and thermal emission (250 μm) after star and galaxy masking and subtraction of the sky. The linear fit using Equation (3) excluded data at high surface brightness in a central rectangular region in Spider containing the brightest cirrus, which corresponds to the plume of black points (low pixel count $n$) that clearly departs from a linear relationship in this figure.

cirrus near the center of Spider was masked using rectangular masks.

At 18″ resolution, there is significant dispersion about the linear correlation line relative to the range of the data. In cases like this, the derived slope of the line is significantly affected by whether the cause of this dispersion is predominantly due to uncertainties in the scattered light measurement or the thermal emission, both of which we have estimated and taken into account in the fit. Nevertheless, as a precautionary mitigation, correlations were performed on images after convolving to 36″, because that drastically reduced the dispersion about the correlation relative to the virtually unchanged range. Figure 3 shows the correlation between Dragonfly (2017 $g$ band) and Herschel 250 μm data after the modeled sky has been subtracted. We found that the derived slope did not change substantially,[15] confirming that our accounting of the uncertainties was satisfactory, but reminding us that there are non-negligible systematic errors too.

The ratios of scattered light to thermal emission calculated for each filter and epoch of Dragonfly data convolved to 36″ are shown in Table 3. The ratios derived for 2014 and 2017 epoch data are consistent with each other. There are systematic errors as well that depend on several different choices during the analysis—relating for example to star removal, modeling of the wide-angle PSF (Liu et al. 2022), order of the polynomial sky component, and angular resolution—and these are of course much more difficult to quantify or to claim to be normally distributed. Our aim in new Dragonfly observations will be to reduce the systematic errors. With these first data, for each passband we simply averaged the results, and took the uncertainty to be half the range of the two values. In summary, the ratio of $g$-band to 250 μm intensity is $\gamma_{g,\,250} = (0.855 \pm 0.025) \times 10^{-3}$ and the ratio of $r$-band to 250 μm is $\gamma_{r,\,250} = (1.55 \pm 0.08) \times 10^{-3}$ when intensities are measured in the same units.

---
[15] An example can be seen in Zhang (2021), Figure 4.13 at 18″ resolution and Figure 4.15 at 36″ resolution, where the slopes determined agree within 5%, which is of the same order as the statistical errors.

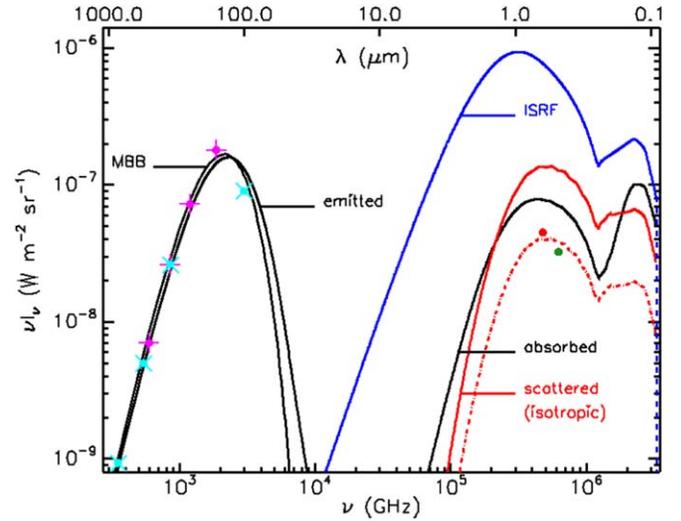

**Figure 4.** The relationship between thermal dust emission and scattered light (see text for details). Data: shape of the thermal emission SED in Spider as measured by Herschel (magenta) and Planck/IRAS (cyan), normalized to 6 MJy sr$^{-1}$ at 250 μm; scattered light (Dragonfly) relative to 250 μm thermal emission (red and green dots). Curves: modified blackbody fit to the data (black); ISRF (blue) cut off at the neutral hydrogen ionization limit (dashed); ISRF times dust model absorption cross section, scaled such that energy absorbed equals energy thermally emitted by dust (black); dust model prediction of the emitted intensity, very close to the MBB (black); ISRF times corresponding scattering cross section in the dust model assuming no anisotropy (red)—the observations are lower than this by about a factor of 0.3 (the level of the dashed–dotted curve).

## 5. Joint Modeling

Figure 4, whose details we will describe in stages, summarizes the principles of how a joint analysis of the thermal dust emission and the scattered light opens opportunities to study both models of interstellar dust and models of the interstellar radiation field.

### 5.1. Data Representation

The first stage was to choose a representative normalization of the Spider intensities, which we take to be 6 MJy sr$^{-1}$ at 250 μm; see the range of Herschel brightness in Figures 2 and 3. This is a magenta plus on the spectral energy distribution (SED) of thermal dust emission in Figure 4, which is on a $\nu I_\nu$ scale. Relative to this we plot the red and green dots for the scattered light from the ratios in Table 3.

The SED was further delineated by correlating Herschel images at the other SPIRE passbands (350 and 500 μm) and at 160 μm from PACS, the Photoconductor Array Camera and Spectrometer (Poglitsch et al. 2010), with that at 250 μm, after bringing the images to the same resolution and grid (magenta pluses). Likewise we incorporated data, again via correlation, from the Planck Legacy Archive[16] at 355, 545, 857 GHz, and 3 THz (cyan crosses), as were used for the Planck all-sky dust model (Planck Collaboration XI 2014).

These measurements of $I_\nu$ of the thermal dust emission were fit with a modified blackbody (MBB) assuming that the dust optical depth $\tau_\nu$ has a frequency dependence described by a power law, with index $\beta = 1.78$, from the Planck dust model. The best fit MBB SED has a dust temperature of about 18 K and $\tau(353\,\text{GHz}) = 7.0 \times 10^{-6}$. It is plotted (in $\nu I_\nu$ form) as a black curve. The integrated intensity in thermal emission, the radiance $\mathcal{R}$ (Martin et al. 2012; Roy et al. 2013; Planck Collaboration XI 2014), is $1.8 \times 10^{-7}$ W m$^{-2}$ sr$^{-1}$, either by numerical integration or from the analytic form for the MBB (Planck Collaboration XI 2014).

---
[16] https://pla.esac.esa.int

### 5.2. Interstellar Radiation Field

Because the MBB represents thermal emission by dust in equilibrium with the local ISRF, then $\mathcal{R}$ is equal to the integral of the intensity of the ISRF times the dust absorption cross section. To calculate this, one needs a description of both ingredients. The commonly used ISRF shown by the blue curve in Figure 4 is from Mezger et al. (1982) as updated by Mathis et al. (1983).

### 5.3. Dust Model of Absorption, Emission, and Scattering

For illustrative purposes, we used the DustEM model from Compiègne et al. (2011) for the diffuse medium at high Galactic latitude, in particular the two components "LamC" and "aSil" (large amorphous carbon and amorphous silicate grains) that are in equilibrium with the above ISRF and produce the thermal dust emission. It is important that for these and similar dust models the size distributions for these components have been constrained by various observations, such as the frequency dependence of the interstellar extinction curve, cosmic abundances, and the relative amount of thermal dust emission.[17]

We calculated the frequency-dependent absorption cross sections for the LamC and aSil size distributions of Compiègne et al. (2011) using Mie theory for homogeneous spherical particles, multiplied their sum by the ISRF, computed the integral of the absorbed radiation, and scaled this to match the observed radiance $\mathcal{R}$. The scaled absorption curve is given in Figure 4 (black). The DustEM code computes the equilibrium dust temperatures and resulting thermal dust emission, the "emitted" SED in Figure 4, which, not surprisingly, is close to the MBB.[18]

Multiplying the scattering cross section by the ISRF using the same scaling gives a prediction of the intensity of scattered light ("scattered" red curve). The observations are a factor about 0.3 (0.5 dex) lower than this (dashed–dotted red curve).[19] However, this prediction assumes that the scattering phase function and/or the ISRF are isotropic. It is important that neither assumption is expected to be valid. For the effects of anisotropy to come into play, both factors need to be anisotropic, as we now discuss.

### 5.4. Anisotropic Phase Function

Because the dust particles are comparable in size to the wavelength, they scatter anisotropically. From Mie theory, we have computed the phase functions for the size distributions for the two compositions and combined them according to the components' contribution to the total scattering cross section. These axisymmetric phase functions are shown for the red and green passbands in Figure 5. There is a strong peak around the forward direction, scattering angle $\theta = 0$. This "forward scattering" can be quantified by the asymmetry parameter $g$, the average of $\cos\theta$ over all directions. For this model, $g$ is 0.55 and 0.57 for red and green passbands, respectively. Also shown is the approximate Henyey–Greenstein phase function (Henyey & Greenstein 1941) for the same values of $g$. It is remarkably similar over the $\theta$ range from the peak to about 125°, but significantly underestimates the amount of backscattering.

---

[17] A small-grain carbonaceous component, SamC, contributes nonequilibrium emission at about the 5% level near the 100 $\mu$m band and starts to contribute to the optical absorption for $1/\lambda > 2$ $\mu$m$^{-1}$.
[18] In detail, the DustEM emission is slightly hotter and broader, because of the LamC contribution to the emission. In a full dust model to be matched to all observations, this would point to a deficiency, but it is unimportant for the purposes here.
[19] Without this lowering, cirrus would be a much larger nuisance for extragalactic observers and much easier to observe for modelers of Galactic dust!

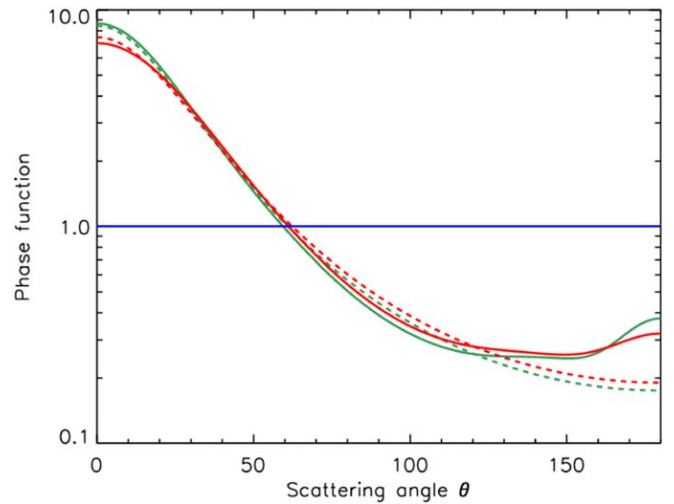

**Figure 5.** Phase functions for the illustrative LamC plus aSil dust model, showing significant reductions relative to isotropic at backscattering angles. Red and green curves are for the corresponding passbands and dashed curves are the Henyey–Greenstein approximations for the same asymmetry parameter.

### 5.5. Anisotropy of the Incident ISRF

Even with an anisotropic phase function, the scattered radiation is isotropic unless the incident radiation making up the ISRF is also anisotropic.

The first such model investigated was that used by Sandage (1976), in which a high-latitude cloud is illuminated by the entire Galactic plane below (hence, IFN), which is taken to be of uniform brightness. As in Sandage (1976), the illumination is axisymmetric and $\Theta$ is the angle between the perpendicular dropped from the cloud to the plane and a line connecting the section of the illuminating plane to the high-latitude cloud. This produces an axisymmetric illuminating intensity proportional to $1/\cos\Theta$. The total illumination is divergent as $-\log\cos\Theta_{\text{max}}$, where $\Theta_{\text{max}}$ effectively defines the extent of the disk, and so Sandage arbitrarily limited $\Theta_{\text{max}}$ to be 89°. Any maximum value lower than 90° can be chosen to explore the changing asymmetry.

We created a second model based on a cloud above a uniform slab that has internal extinction characterized by optical depth $\tau$ from the midplane vertically to the edge of the slab. The intensity is then proportional to $1 - \exp(-2\tau/\cos\Theta)$. This follows the same equations as the model used by Mezger et al. (1982) to evaluate the intensity seen by an observer at the midplane of a uniform slab, except for the factor 2 (because the cloud is outside the slab), and we adopt their scale of $\tau$, adjusting for red and green passbands using the CCM extinction curve (Cardelli et al. 1989) for the ratio of total to selective extinction $R_V = 3.1$. Values of $\Theta_{max}$ can range up to 90°, because the extinction removes the divergence.

Third, we used the ISRF models from the FRaNKIE code, a tool for calculating multiwavelength interstellar emissions in galaxies (Porter & Strong 2005; Porter et al. 2015, 2017). These are part of the GALPROP code for cosmic-ray transport and diffuse emission production and were downloaded from that site.[20] Based on a model of the Galaxy, these are more realistic than the first two models. The cloud is illuminated from all directions and the illumination changes with the height of the cloud above the Galactic plane and Galactocentric distance. The most appropriate for our analysis of the Spider observations are the FRaNKIE models for the ISRF at the Galactocentric distance of 8.5 kpc and 0.2 (and 0.5) kpc above the Galactic plane, evaluated for filters at 0.6 and 0.45 μm. There is not only an asymmetry with Θ (or Galactic latitude), but also an azimuthal asymmetry (with Galactic longitude) because the illumination from directions in the inner Galaxy is higher.[21] In practice, the illuminating intensity is given on a HEALPix sphere (Górski et al. 2005), with nside = 32 (12,288 pixels).[22]

### 5.6. The Combined Effects of Anisotropy

Using all-sky pixelated descriptions of the illumination of dust in the Spider, and the phase functions, we calculate pixel by pixel the resulting intensity of radiation scattered in any particular direction, including toward us. Summing this up and comparing it to the all-sky average of the radiation scattered, we get an estimate of the ratio of the scattered intensity in any chosen direction to the average, the "isotropic" curve in Figure 4.

The ratio can be larger or smaller than unity, depending on the direction: the geometry is important. Given the particular geometrical arrangement of the Spider, us, and plausible models of the orientation of the ISRF anisotropy, there is a lot of sideways scattering to backscattering in all models, and so also given the plausible phase functions from the interstellar dust models, the ratio for the scattering direction toward us is less than unity. This is qualitatively in the correct sense to explain the observations being lower than the isotropic curve, demonstrating the importance of the anisotropies. We report quantitative values of the ratio for the red (green) passbands below.

For the Sandage model the ratio is 0.59 (0.56) for $\Theta_{max} = 89°$, decreasing to 0.50 (0.47) and 0.37 (0.35) for 85° and 70°, respectively, as the IRSF anisotropy increases. The ratios are a bit lower for the green passband because of the more forward-scattering phase function.

For these same three angles, the ratios for the uniform slab model are 0.50, 0.46, and 0.37 (0.46, 0.42, and 0.34). Here the radiation field is not so highly anisotropic at high $\Theta_{max}$ and there is a small extra effect because $\tau$ is higher in the green passband. The different limits on $\Theta_{max}$ are not well motivated and both of these models are unphysically axisymmetric. Nevertheless, they give some sense of the sensitivity to plausible anisotropies.

For the red (green) FRaNKIE model, the ratio is 0.63 (0.64) for height 0.2 kpc, decreasing to 0.51 (0.49) for height 0.5 kpc because the anisotropy of the ISRF increases. We experimented with median-filtering the IRSF angular distribution to mitigate the effects of Monte Carlo noise spikes.[23] The ratios became 0.56 (0.53) and 0.44 (0.41), revealing some sensitivity.

### 5.7. Discussion

This quantitative analysis shows the importance of modeling the anisotropies for fully understanding the ratio of scattered light to thermally emitted radiation. However, none of these ratios for various representations of the ISRF is as small as the factor of 0.3 indicated by the modeling in Figure 4.

Lowering the albedo could affect the joint modeling result, by lowering the isotropic curve. However, the albedo cannot be changed at will. The albedo depends on the complex refractive index of the different dust grain materials and the size distributions, and these are already constrained by other observations in the dust modeling, of which the DustEM model used is one representative example. Nevertheless, it might be a factor contributing to a lower ratio and better quantitative agreement. Another somewhat related factor is that the energy absorbed by the dust involves the full spectrum of the ISRF and the absorption cross sections across this spectrum, whereas, for the time being, we do not have a full spectrum of the scattered light.

Alternative dust models could be explored as well, such as that of Draine & Li (2007) and its predecessors. However, it is a two-component model similar to DustEM and there is a basic problem that these models underpredict the ratio of thermal emission to optical extinction by about a factor of two (Planck Collaboration Int. XXIX 2016). Furthermore there is a similar problem, with the predicted polarized thermal emission at 353 GHz (Draine & Fraisse 2009) being too low relative to optical interstellar polarization (Planck Collaboration Int. XXI 2015). Guillet et al. (2018) have proposed modifications of the DustEM model, in grain shape and alignment and introducing the possibilities of porosity and carbonaceous inclusions on the silicate component. A further challenge for two-component dust models is the relatively flat spectral dependence of the fractional polarization at the submillimeter wavaelengths seen by Planck and BLASTpol (Planck Collaboration Int. XXII 2015; Ashton et al. 2018). A composite single-component model addresses this (Draine & Hensley 2021; Hensley & Draine 2022).

We have not worked out the consequences of these new models. Favorable features to look for would be an even more anisotropic phase function and a lower albedo.

There is an elevated level of the observed scattered light in the red passband compared to the green, not accounted for by any of the combined models. Perhaps this is an indication of a contribution from the "extended red emission" (ERE) phenomenon (Witt et al. 2008), but it would need to be small because ERE is excited by near-UV photons and is isotropic, not subject to the phase function.

---

[20] https://galprop.stanford.edu/, specifically the data file for GALPROP version 54 (and 56), galprop-54_data_081715.tar.gz
[21] It would be possible to add an azimuthal asymmetry to the Sandage model and the uniform disk model, for example by considering a radially exponential disk.
[22] https://healpix.sourceforge.io

[23] The radiation fields are computed by Monte Carlo techniques and are not smooth on fine angular scales.

As dust models and ISRF models are being refined it will be helpful to have more data for analysis such as presented in this paper to find what will resolve the discrepancy. One promising IFN observed by Witt et al. (2008) that we have imaged with Dragonfly is the Draco nebula (Miville-Deschênes et al. 2017). It shows a strong linear relationship between scattered light and Herschel submillimeter thermal dust emission, at least in the part of the cloud that is not molecular (to be optically thin). The Draco field is centered on $\ell \sim 91°$, $b \sim 38°$ and is at a distance of about 600 pc, thus at a height $z = 370$ pc above the Galactic plane and, like the Spider, outside the Local Bubble. Being at a different Galactic position, the scattering geometry could provide a distinctively different coupling to the anisotropy of the ISRF.

## 6. Near-infrared to Far-ultraviolet (FUV) Observations of the Ratio of the Scattered DGL to Thermal Dust Emission at 100 μm

Traditionally, and even recently, the intensity of scattered light has been compared to and/or extracted by correlation with dust emission at 100 μm. The latter passband is past the peak of the SED and so is not the best tracer of the column density of dust because of the temperature sensitivity. It is now preferable to use submillimeter emission, as captured in the all-sky Planck thermal dust model (Planck Collaboration XI 2014), which has about the same spatial resolution as IRAS, 5′. Herschel has submillimeter passbands and at 36″ about 10 times the resolution, so that our correlation diagrams are not sparse as in early work on high-latitude clouds (IFNs).

Nevertheless, for comparison our results can be rescaled to be relative to 100 μm emission by multiplying $\gamma_{g,250}$ and $\gamma_{r,250}$ by $I_{250}/I_{100} = 2$, the value typical of the Spider emission, to obtain $\gamma_{g,100} = (1.71 \pm 0.05) \times 10^{-3}$ and $\gamma_{r,100} = (3.10 \pm 0.16) \times 10^{-3}$ for the two optical filters. These are in the same range as seen in the comprehensive compilation of earlier results of scattered light correlated with 100 μm emission (Figure 6 and Table 3) in Ienaka et al. (2013), but this is probably not particularly instructive given that these "optical data are a collection of heterogeneous samples that were taken with different techniques and analyzed using different methods," including differential measures in targeted observations of clouds (Witt et al. 2008) and low-resolution scanned DGL observations of the high-latitude sky by Pioneer 10/11 (Matsuoka et al. 2011). It was not always possible to avoid the effects of optical depth explicitly, which is of paramount importance for our investigation.

Another approach to the DGL exploits archival blank-sky spectra in SDSS-II (Brandt & Draine 2012), Hubble Space Telescope Faint Object Spectrograph (HST/FOS; Kawara et al. 2017), and SDSS-III/BOSS (Chellew et al. 2022), which, even though they are poorly matched in spatial resolution to 100 μm imaging data, have the statistical power to reveal the DGL correlation spectrum. These are expressed as the ratio $\alpha = (\nu I_\nu)_{\rm opt}/(\nu I_\nu)_{100}$. To convert our $\gamma_{g,250}$ and $\gamma_{r,250}$ to this convention, using $\lambda_0$ for the filters from Table 1 we multiply by $(100/\lambda_0)I_{250}/I_{100}$ to obtain $\alpha_{g,100} = 0.36 \pm 0.01$ and $\alpha_{r,100} = 0.50 \pm 0.03$ for the $g$ and $r$ filters, respectively. These broadband filter values are quite close to the respective values in the $C\alpha'$ correlation spectrum in the left panel of Figure 7 in Chellew et al. (2022) for the BOSS northern sky, which incorporates their bias factor $C = 2.1$ and correction (denoted by the prime) to remove the dependence on optical

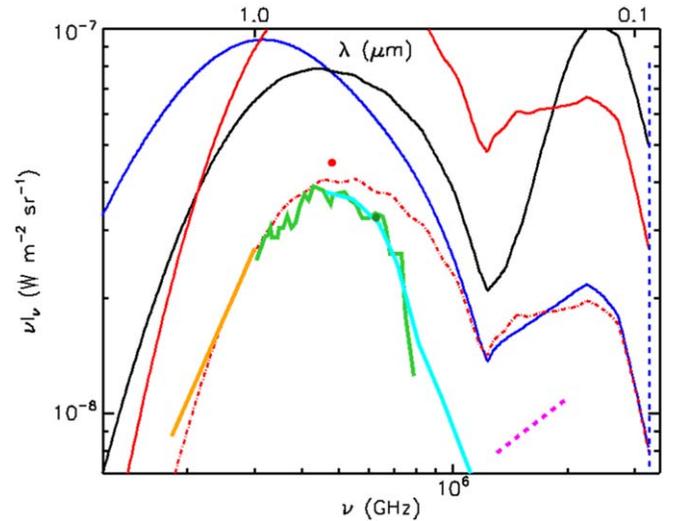

**Figure 6.** Zoom to highlight the near-infrared to near/far-ultraviolet portion of Figure 4. To provide the context of the frequency dependence of the ISRF (blue), that curve has been multiplied by 0.1. New additions are overplots of our coarse sampling of measurements of correlation spectra of scattered DGL relative to 100 μm: BOSS spectrum (Chellew et al. 2022) in lime green; CIBER spectrum (Arai et al. 2015), renormalized to match the BOSS spectrum at 1 μm (orange); FOS spectrum (Kawara et al. 2017) with the same renormalization (cyan); likewise, two Galaxy Evolution Explorer (GALEX) passbands (joined by a dashed magenta line) from Murthy et al. (2010). See text for details and discussion. The upper x-axis ranges from 2.5 μm on the left to 0.0857 μm on the right.

depth (corrected spectra are very close to those in the optically thin limit).

However, compared to the scattered light from a distinctive isolated cloud (IFN) that is also optically thin, as we have studied, the DGL is conceptually quite different and is traditionally modeled with Galactic distributions of dust and stars that are described by smooth functions of height above the Galactic disk. Recent descriptions of the Local Bubble in the solar neighborhood (Lallement et al. 2019; Leike et al. 2020; Pelgrims et al. 2020; Zucker et al. 2022; Marchal & Martin 2023) indicate that a more sophisticated treatment of the DGL in the high-latitude sky will be required. Still, the DGL is some averaged response to the anisotropies discussed over a considerable part of the high-latitude sky. By studying individual clouds (IFNs), it is in principle possible to distinguish more clearly the dependence from geometry coupled with the anisotropies (Section 5.7).

Nevertheless, to get a sense of the frequency dependence of the scattered DGL, beyond the $g$ and $r$ passbands, we plotted this $C\alpha'$ correlation spectrum in Figure 6 (green line), having multiplied it by the $\nu I_\nu$ value at 100 μm in Figure 4, $8.96 \times 10^{-8}$ W m$^{-2}$ sr$^{-1}$, to get the appropriate normalization. This spectrum happens to pass through our $g$-filter measurement.

Similarly, we plotted the trend of the correlation spectrum in the near-infrared found in the sounding rocket observations of the Cosmic Infrared Background Experiment (CIBER) in the range from 0.95 to 1.65 μm (Arai et al. 2015). For comparison purposes, it is common to renormalize DGL observations to account for the different sky coverage, and here we did the same, multiplying the CIBER results by a factor 0.6 to match the $C\alpha'$ spectrum at 1 μm (red line). The Cosmic Background Explorer Diffuse Infrared Background Experiment (COBE/DIRBE) value at 1.25 μm is lower, but compatible within its uncertainty (Sano et al. 2015). We do not show any results for wavelengths longer than 2 μm, because there the correlated DGL has a component related to nonequilibrium emission by very small dust particles or large molecules; see the summary Figure 5 in Sano et al. (2016).

We also plotted the FOS correlation spectrum to extend the data to 0.27 μm in the near-ultraviolet (NUV; Kawara et al. 2017). We multiplied the FOS spectrum by the same factor used to rescale CIBER, and it can be seen that this provides a good match to the $C\alpha'$ spectrum in the optical (blue line). This good match could be anticipated from the compilation in Figure 9 by Kawara et al. (2017). To extend to higher frequencies, we plotted the spectral dependence of the correlated emission derived from the two broad passbands from GALEX (NUV at 0.231 μm and FUV at 0.153 μm) by Murthy et al. (2010) for optical depths less than 0.7, again with the same scaling (dashed purple line), though this might be questioned. The all-sky ratio for GALEX FUV found by Hamden et al. (2013) is 7% lower, but the value found by Seon et al. (2011) for the L-band range of the Spectroscopy of Plasma Evolution from Astrophysical Radiation (SPEAR, also known as FIMS) spectrograph, which is the same as the GALEX FUV channel, is about a factor of two lower.

This compiled correlation spectrum is not necessarily what would be observed in the Spider. However, there are interesting general points to note for perspective on modeling. Toward the near-infrared the observed DGL continues to fall, despite the probability that the scattering becomes less anisotropic. In optically thin integrated flux nebulae like the Spider, it would be very difficult to isolate the faint scattered light in the near infrared. Two main factors are the decrease in the ISRF and the steep decrease in the scattering cross section. The latter is particularly sensitive to the large-particle end of the size distribution (Brandt & Draine 2012). By comparison, the peak in scattered light in the optical is a sweet spot of lower sensitivity to these factors, because the product of the ISRF and the scattering cross section has a broad peak.

One complication in the red part of the optical spectrum is the ERE, which is thought to be excited by near-UV photons and is emitted isotropically even if the ISRF is anisotropic, in contrast to the scattered light detected, which is suppressed by the anisotropic phase function. By comparison to simple DGL models, Chellew et al. (2022) found evidence for an ERE excess at 0.65 μm, stronger in the southern Galactic hemisphere. This falls in the red Dragonfly band, and so for that band it would frustrate the simplest joint modeling we have presented; the green band would not be affected.

Toward the ultraviolet, the scattered light falls because of three main factors—the decreasing ISRF (including the observed 4000 Å break), the flattening of the scattering cross section of the particles in the large grain size distribution, and increased anisotropy of the scattering (given the probable geometry). The FOS results (Kawara et al. 2017) indicate a continued decline, to at least 0.27 μm ($1.11 \times 10^6$ GHz). There is an additional matter of extinction increasing with frequency so that the Spider would no longer be in the optically thin limit, further diminishing the scattered light ratio.

In the GALEX range, the DGL correlation spectrum shows a rise. A significant factor would be the increase in the IRSF, contributions from hot stars. In fact, illumination by individual stars becomes important (Henry 1977; Henry et al. 1977; Seon et al. 2011; Hamden et al. 2013; Murthy 2016), so that there are significant local excursions relative to the average DGL. This range is certainly not relevant for simple joint modeling of the Spider, because of the opacity. The Spider and the entire NCPL are notably faint in the all-sky maps of the ratio of scattered light to 100 μm emission; see Figure 4 in Murthy et al. (2010). Furthermore, modeling the ultraviolet behavior eventually brings in scattering by small-grain components with nonequilibrium emission (and with scattering that is less anisotropic than for the large particles) and so diverges from the assumptions of the joint modeling that we have presented.

## 7. Summary and Conclusion

In this paper, we quantified the relationship between scattered light and thermal emission from a diffuse cirrus cloud, Spider, using data from Dragonfly and Herschel.

A key challenge for optical observations of faint cirrus is accurate removal of a spatially varying background. Via jackknife tests, we demonstrate that correlating pixel values of two independent observations of the same cirrus field while simultaneously fitting the sky-related signal as a noncorrelating additive component is a robust method (Section 3.3).

This method of concurrent correlation of two images and modeling of the sky-related signal was applied to Dragonfly g- and r-band images to determine the color of Spider, which we measured to be $g - r = 0.644 \pm 0.024$ (Section 3.4). Our color measurement is consistent with Román et al. (2020), who measured the $g - r$ color of Galactic cirri in the Stripe82 region using Sloan Digital Sky Survey data reprocessed by modeling and removing the instrument scattered light produced by stars; the $g - r$ color was in the range 0.5 to 0.9, with a median of 0.63.

The method was applied again to the combined analysis of Dragonfly and Herschel images of Spider. We measured the ratio of the g- and r-band scattered light intensities to the 250 μm thermal dust emission intensity to be $\gamma_{g,250} = (0.855 \pm 0.025) \times 10^{-3}$ and $\gamma_{r,250} = (1.55 \pm 0.08) \times 10^{-3}$, respectively, when the intensities are measured in the same units (Section 4).

We demonstrated how observed thermal emission combined with any dust model and ISRF model can be used to predict the amount of scattered light when the cloud is optically thin (Section 5). Comparing this prediction with the observed intensity of scattered light is a means to test a combination of dust and ISRF models. Starting from the measured thermal emission and using the DustEM model from Compiègne et al. (2011), the measured scattered light intensity is 0.3 times fainter than the predicted value would be when assuming an isotropic ISRF and/or an isotropic dust scattering phase function. However, the ISRF coming from the Galaxy is backscattered into our line of sight by Spider. We considered different models of the dust and ISRF, but none of these models lowers the predicted scattered light intensity by as much as a factor of 0.3. To resolve this discrepancy, dust models and ISRF models will need to be refined, and it will be helpful to carry out the joint analysis on other isolated clouds (IFNs) that provide a distinctively different coupling to the anisotropy of the ISRF (Section 5.7).

Further distinctive diagnostics of the dust and ISRF could be provided by measurements of the fractional polarization and the polarization angle of the scattered light (S. K. Bowes & P. G. Martin 2022, in preparation).

Even in the absence of perfected models of the dust and ISRF, the demonstration of a strong correlation of scattered light and thermal emission opens up an empirical approach in which any particular deep wide-field optical observations can be correlated with the Planck all-sky thermal dust emission (Planck Collaboration XI 2014) to quantify locally where low-level scattered light contamination would be present. Scattered light can reveal the spatial structure of dust at much higher resolution.

We are very grateful to the staff at New Mexico Skies Observatories, without whom this work could not have been carried out. We acknowledge support from the Natural Sciences and Engineering Research Council (NSERC) of Canada. The work done by J.Z. for this paper was supported through grants from the Ontario government. This research has made use of the NASA Astrophysics Data System, the variable star observations from the AAVSO International Database contributed by observers worldwide, and observations obtained with Planck (http://www.esa.int/Planck). Planck is an ESA science mission with instruments and contributions directly funded by ESA Member States, NASA, and Canada.

The Herschel spacecraft was designed, built, tested, and launched under a contract to ESA managed by the Herschel/Planck Project team by an industrial consortium under the overall responsibility of the prime contractor Thales Alenia Space (Cannes), and including Astrium (Friedrichshafen) responsible for the payload module and for system testing at spacecraft level, Thales Alenia Space (Turin) responsible for the service module, and Astrium (Toulouse) responsible for the telescope, with in excess of a hundred subcontractors.

PACS has been developed by a consortium of institutes led by MPE (Germany) and including UVIE (Austria); KU Leuven, CSL, IMEC (Belgium); CEA, LAM (France); MPIA (Germany); INAF-IFSI/OAA/OAP/OAT, LENS, SISSA (Italy); IAC (Spain). This development has been supported by the funding agencies BMVIT (Austria), ESA−PRODEX (Belgium), CEA/CNES (France), DLR (Germany), ASI/INAF (Italy), and CICYT/MCYT (Spain).

SPIRE has been developed by a consortium of institutes led by Cardiff University (UK) and including Univ. Lethbridge (Canada); NAOC (China); CEA, LAM (France); IFSI, Univ. Padua (Italy); IAC (Spain); Stockholm Observatory (Sweden); Imperial College London, RAL, UCL−MSSL, UKATC, Univ. Sussex (UK); and Caltech, JPL, NHSC, Univ. Colorado (USA). This development has been supported by national funding agencies: CSA (Canada); NAOC (China); CEA, CNES, CNRS (France); ASI (Italy); MCINN (Spain); SNSB (Sweden); STFC, UKSA (UK); and NASA (USA).

We thank the anonymous referee whose comments and suggestions have improved this manuscript.

*Facilities:* Dragonfly Telephoto Array, Herschel, Planck, IRAS, AAVSO.

*Software:* Astrometry.net (Lang et al. 2010), astropy (Astropy Collaboration et al. 2013), SExtractor (Bertin & Arnouts 1996), SkyMaker (Bertin 2009), and SWarp (Bertin et al. 2002).

### ORCID iDs

Jielai Zhang (张洁莱) https://orcid.org/0000-0001-5310-4186
Peter G. Martin https://orcid.org/0000-0002-5236-3896
Ryan Cloutier https://orcid.org/0000-0001-5383-9393
Natalie Price-Jones https://orcid.org/0000-0002-4220-1682
Roberto Abraham https://orcid.org/0000-0002-4542-921X
Pieter van Dokkum https://orcid.org/0000-0002-8282-9888
Allison Merritt https://orcid.org/0000-0001-9467-7298